\documentclass{twocol}
\usepackage{flushend}
\twocolumn

\lat

\title{
Linear NMR in the polar phase of $^3$He in aerogel
}

\rtitle{Linear NMR in the polar phase of $^3$He in aerogel
}

\sodtitle{
Linear NMR in the polar phase of $^3$He in aerogel
}

\author{
V.~V.~Zavjalov\/\thanks{e-mail: vladislav.zavyalov@aalto.fi},
}

\address{Low Temperature Laboratory,
Department of Applied Physics, Aalto University,
PO Box 15100, FI-00076 AALTO, Finland}

\rauthor{V.~V.~Zavjalov}

\sodauthor{Zavjalov}

\abstract{$^3$He is an example of the system with non-trivial Cooper
paring. A few different superfluid phases are known in this system.
Recently the new one, the polar phase, have been observed in $^3$He
confined in nematically ordered aerogel. A number of various topological
defects including half-quantum vortices can exist the polar phase. In
this work we present theoretical and numerical studies of linear NMR in
the polar phase both in the uniform order-parameter texture and in the
presence of half-quantum vortices. }


\graphicspath{{pics/}}
\newcommand{\image}[3]{
\begin{figure}[#1]
\begin{center}
\includegraphics{full_#2.eps}
\caption{\small#3}
\label{image:#2}
\end{center}
\end{figure}
}
\newcommand{\imagew}[3]{
\begin{figure*}[#1]
\begin{center}
\includegraphics{full_#2.eps}
\caption{\small#3}
\label{image:#2}
\end{center}
\end{figure*}
}

\begin{document}

\maketitle

\subsection*{Introduction}

The polar phase of~$^3$He in nematically ordered aerogel has been
predicted in~\cite{2006_aoyama} and found experimentally
in~\cite{2015_dmitriev_polar}. In this system a new topological defect, a
half-quantum vortex can exist. Half-quantum vortices have been originally
predicted for A-phase in \cite{1976_volovik_mineev} but have not been
observed in experiments. This is because of energetically unfavorable
solitons which should always connect half-quantum vortex pairs in the
A-phase. In the polar phase of~$^3$He in aerogel there are no such solitons if the
magnetic field is parallel to aerogel strands. Also vortices in the polar
phase are strongly pinned and cannot move when the field is tilted and
solitons appear. In our experimental work~\cite{2015_polar_rota} vortices were
created by rotating the~$^3$He sample. Then the field was tilted and
spin waves localized in solitons were observed by NMR. In this paper we
develop a theory for textures, topological defects and spin dynamics in
the polar phase of $^3$He. We also do numerical simulations of spin waves
in the presence of half-quantum vortices.

\vbox{
\subsection*{Order-parameter and energies}}

We are studying the polar phase of~$^3$He in nematically ordered aerogel.
The order parameter in this system (\cite{2016_mineev}) is
\begin{equation}\label{eq:order_par}
A_{aj}  = \frac{1}{\sqrt{3}}\Delta\ e^{i\varphi} d_a l_j,
\end{equation}
where~$\varphi$ is the phase, and $\bf d$ and $\bf l$ are unit vectors in
spin and orbital spaces respectively. The orbital unit vector~$\bf l$ is directed
along the aerogel strands and can not move.

There are three components of the Hamiltonian which are important for
spin dynamics: magnetic energy, energy of spin-orbit interaction and
gradient energy:
\begin{equation}\label{eq:ham}
\mathcal{H} = F_M + F_{SO} + F_\nabla,
\end{equation}
\begin{eqnarray}
\label{eq:en_m0}
F_M &=& - ({\bf S} \cdot \gamma {\bf H})
+ \frac{\gamma^2}{2}\chi_{ab}^{-1} S_a S_b,\\
\label{eq:en_d0}
F_{SO}
&=& 3g_D \Big[
           A^*_{jj}A_{kk}
         + A^*_{jk}A_{kj}
 - \frac23 A^*_{jk}A_{jk}\Big],\\
\label{eq:en_g0}
F_\nabla
&=& \frac32 \Big[
    K_1 (\nabla_j A^*_{ak})(\nabla_j A_{ak})\\\nonumber
&+& K_2 (\nabla_j A^*_{ak})(\nabla_k A_{aj})
+   K_3 (\nabla_j A^*_{aj})(\nabla_k A_{ak}) \Big],
\end{eqnarray}
where~$\bf S$ is spin and $\bf H$ is the magnetic field.
Susceptibility~$\chi_{ab}$ is anisotropic, the axis of anisotropy is~$\bf d$
and minimum of the magnetic energy corresponds to $\bf S \perp d$. This
can be written as
\begin{equation}
\chi_{ab}^{-1} =
\frac{1}{\chi_\perp} ( \delta_{ab} + \delta\ d_a d_b)
,\quad
\delta = (\chi_\perp-\chi_\parallel)/\chi_\parallel > 0.
\end{equation}

Substituting the order parameter~(\ref{eq:order_par}) into energies and
using the fact that~$\bf l$ is uniform we have
\begin{eqnarray}
\label{eq:en_m1}
F_M &=& - ({\bf S} \cdot \gamma {\bf H})
+ \frac{\gamma^2}{2\chi_\perp}
\left[{\bf S}^2 + \delta\ ({\bf d} \cdot {\bf S})^2\right],\\
\label{eq:en_d1}
F_{SO}
&=& 2\Delta^2g_D\ \Big[({\bf d} \cdot {\bf l})^2-\frac13\Big],\\
\label{eq:en_g1}
F_\nabla
&=& \frac{\Delta^2}{2} K_{jk}\ [
  (\nabla_j \varphi) (\nabla_k \varphi) + (\nabla_j d_a)(\nabla_k d_a)],
\end{eqnarray}
where symmetric matrix~$K_{jk} = K_1 \delta_{jk} + (K_2+K_3) l_j l_k$ is
introduced. Motion of the phase~$\varphi$ (sound) is not coupled with the
motion of~$\bf d$ (spin waves). In spin dynamics terms with the phase
gradients give only a constant contribution to the energy and can be
skipped.


\subsection*{Equilibrium texture}

Let's first study the static picture. In the equilibrium
$\partial\mathcal{H}/\partial S_a=0$. This means
\begin{equation}\label{eq:dHdS}
{\bf S^0} + \delta\ ({\bf d^0} \cdot {\bf S^0})\ {\bf d^0}
=\frac{\chi_\perp}{\gamma}\ {\bf H},
\end{equation}
where~$\bf S^0$ and~$\bf d^0$ are equilibrium values of~$\bf S$ and~$\bf
d$. Multiplying this by~${\bf d^0}$ we can find $({\bf d^0} \cdot
{\bf S^0}) =\chi_\parallel/\gamma \ ({\bf d^0} \cdot {\bf H})$. then substituting
it back to~(\ref{eq:dHdS}) we find the value for the spin in the equilibrium:
\begin{equation}
\gamma S^0_a =
\Big[ \chi_\perp \delta_{ab} - (\chi_\perp - \chi_\parallel) d^0_a d^0_b\Big] H_b
= \chi_{ab} H_b
\end{equation}

For calculation of the equilibrium distribution (texture) of the~$\bf d$
vector we will use a coordinate system where~$\bf H\parallel\bf\hat z$
and~$\bf l$ is in~$\bf\hat z- \hat y$ plane (See Fig.1). This can be written as
\begin{eqnarray}\label{eq:angles}
&&{\bf H} = {\bf\hat z} H
,\qquad
{\bf l} = {\bf\hat y} \sin\mu + {\bf\hat z} \cos\mu
,\\\nonumber
&&{\bf d^0} = ({\bf\hat x} \cos\alpha
 + {\bf\hat y} \sin\alpha)\sin\beta + {\bf\hat z}\cos\beta.
\end{eqnarray}
Here~$\mu$ is angle between~$\bf l$ and magnetic field, it is set by the
experimental setup because direction of~$\bf l$ is determined by
aerogel; $\beta$ is angle between~$\bf d^0$ and the field; $\alpha$ is
azimuthal angle of~$\bf d$ in the plane, perpendicular to the
magnetic field, it is counted from the line, perpendicular to both~$\bf
H$ and~$\bf l$ which corresponds to the minimum of energy.

\image{h}{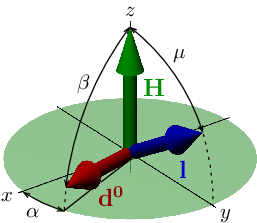}{Fig.~1. Angles, used in the texture calculations}

The energies~(\ref{eq:en_m1})-(\ref{eq:en_g1}) (without constant terms) are:
\begin{eqnarray}
\label{eq:en_m2}
F_M &=&
\frac12(\chi_\perp - \chi_\parallel)H^2 \ \cos^2\beta,
\\
\label{eq:en_d2}
F_{SO}
&=& 2g_D\Delta^2
\ (\sin\alpha\sin\beta \sin\mu + \cos\beta \cos\mu)^2,
\\
\label{eq:en_g2}
F_\nabla
&=& \frac{\Delta^2}{2} K_{jk}
\ [\sin^2\beta (\nabla_j\alpha)(\nabla_k\alpha) + (\nabla_j\beta)(\nabla_k\beta)]
\end{eqnarray}

There are two scales introduced by these energies. Ratio of magnetic and
gradient energies gives the magnetic length~$\xi_H$ and ratio of
spin-orbit and gradient energies gives the dipolar length~$\xi_D$. Since
the gradient energy is anisotropic, we have different values in
directions perpendicular and parallel to the~$\bf l$ vector:
\begin{equation}
\xi^2_{Hjk} = \frac{K_{jk}\Delta^2}{H^2 (\chi_\perp - \chi_\parallel)}
,\quad
\xi^2_{Djk} = \frac{K_{jk}}{4 g_D}
\end{equation}
In the high-field limit $\xi_D\gg\xi_H$. Magnetic energy is in the
minimum everywhere excluding small regions of the~$\xi_H$ size (for
example cores of spin vortices). The small volume of this regions makes
them invisible in NMR experiments. In the rest of the
volume~$\beta=\pi/2$, only variations of~$\alpha$ are important and the
energy is:
\begin{equation}
\label{eq:en_alpha}
\mathcal{H}
= \frac12 K_{jk}\Delta^2
\ (\nabla_j\alpha)(\nabla_k\alpha)
+ 2g_D\Delta^2\ \sin^2\alpha\sin^2\mu
\end{equation}

The equilibrium state corresponds to the
minimum:~$\delta\mathcal{H}/\delta\alpha = 0$. Since the energy depends
on the gradient we have to use variational derivative
\begin{equation}
\frac{\delta\mathcal{H}}{\delta\alpha}
=
\frac{\partial\mathcal{H}}{\partial\alpha}
- \nabla_j\frac{\partial\mathcal{H}}{\partial\nabla_j\alpha}.
\end{equation}
Using this for energy~(\ref{eq:en_alpha}) we have a simple equation
for the distribution of~$\alpha$:
\begin{eqnarray}
\label{eq:eq_alpha}
\bar\xi^2_{jk}\ \nabla_j\nabla_k\alpha
&=& \frac12 \sin2\alpha
,\\\nonumber
\qquad\mbox{where}\quad
\bar\xi_{jk}
&=& \frac{\xi_{Djk}}{\sin\mu}.
\end{eqnarray}
One can see that in the case of~$\bf H\parallel \l$ (or $\mu=0$) there
is no length scale in this problem. $\bf d$ can freely move in the plane
perpendicular to the field and only the gradient term is important.
Tilting the magnetic field from the~$\bf l$ direction makes
the length~$\bar\xi$ finite. At~$\bf H\perp l$ the length scale reaches
its minimum value,~$\xi_D$.


\subsection*{Textural defects}

Equation~(\ref{eq:eq_alpha}) shows that in a tilted magnetic field there
are two possible uniform textures with~$\alpha=0$ and~$\alpha=\pi$.
Vector~$\bf d$ is oriented perpendicularly to both~$\bf H$ and~$\bf l$
and can point in two possible directions. Between this two states there
is a {\it $d$-soliton}. One can also imagine a {\it spin vortex} in which
vector $\bf d$ rotates by $2\pi$ around the vortex line. Two $d$-solitons
should end at this vortex. Looking at the order parameter
formula~(\ref{eq:order_par}) one can see that there can be also a {\it
half-quantum vortex}, in which both vector~$\bf d$ and phase~$\phi$
rotate by~$\pi$ around the vortex line. This is possible because
$A_{\alpha j}({\bf d}, \phi) = A_{\alpha j}(-{\bf d}, \phi+\pi)$. In the
tilted magnetic field one $d$-soliton should end at the half-quantum
vortex. On Fig.~2 two types of vortices are shown.

\image{h}{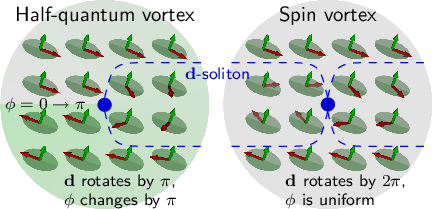}{Fig.~2. The half-quantum vortex and the spin vortex
in the polar phase of $^3$He. Vector~$l$ is perpendicular to the picture
plane. Angle~$\alpha=0$ is changing by~$\pi$ between upper and lower
parts of the picture. This can be done via either a $d$-soliton or a
$\pi$ jump in the phase (which is shown by color gradient).}

The form of the single $d$-soliton can be found analytically.
In this one-dimentional problem equation~(\ref{eq:eq_alpha}) has a form of
static sine-Gordon equation:
\begin{equation}\label{eq:d2alpha}
\bar\xi^2\ \alpha''(x) = \frac12 \sin2\alpha(x),
\end{equation}
where $x$ is a coordinate perpendicular to the soliton. Here the value
of~$\bar\xi$ depends on the soliton orientation: if~$x$ coordinate goes
perpendicular or parallel to~$\bf l$, it should be~$\bar\xi_{\perp}$
or~$\bar\xi_{\parallel}$ respectively.

The analytical solution can be obtained by multiplying the equation by
$a'$ and integrating with proper boundary conditions. Then for a single soliton
with $\sin\alpha(\pm\infty)=0$ and $\alpha'(\pm\infty)=0$ we have
\begin{equation}\label{eq:dalpha2}
\bar\xi^2\ (\alpha')^2 = \sin^2\alpha,
\end{equation}
and then for the soliton located at $x=0$:
\begin{equation}\label{eq:alpha_sol}
\alpha(x) = 2\arctan\left(\exp(x/\bar\xi)\right)
\end{equation}

In the 2D case with isotropic $\xi_D$ (which takes place when the texture
is uniform along $\bf l$-direction) the sine-Gordon equation has analytic
solutions for a number of configurations with spin vortices and solitons
\cite{Hudak1982,Nakamura1983}. This includes, in particular, the kink on
soliton, which  represents the one-quantum ($2\pi$) spin vortex with
two $d$-solitons being on the opposite sides of it (see right part of
Fig.~2). The linear chain of the alternating $2\pi$ and $-2\pi$ vortices
has also analytic solution. The configuration  with two solitons crossing
each other may also represent the spin vortex, if each soliton has a kink
and the positions of two kinks coincide. This is $4\pi$ spin vortex, from
which four $d$-solitons emerge. Such analytic solutions do not take into
account the pinning of vortices which exists in the real system.


\subsection*{Spin dynamics}

To study spin dynamics we write Hamilton equations using Poisson brackets.
Motion of any value $A$ in this approach is given by $\dot A =
\{\mathcal{H},A\}$. Choice of coordinates is quite arbitrary as far as we know
Poisson brackets for them. Brackets can be found from microscopic
considerations, from commutation rules in quantum mechanic, or from
symmetry~\cite{poisson}. For spin~$\bf S$ and a vector~$\bf d$ in the spin space the
Poisson brackets are
\begin{equation}\label{eq:brackets_d}
\{S_a, S_b\} = -e_{abc} S_c, \quad
\{d_a, d_b\} = 0,
\end{equation}
$$
\{d_a, S_b\} = \{S_a, d_b\} = -e_{abc} d_c,
$$
and equations of motion:
\begin{eqnarray}\label{eq:ham_eq0a}
\dot S_a
= \{\mathcal{H}, S_a\}
&=&
  \frac{\delta \mathcal{H}}{\delta S_b} \{S_b,S_a\}
+ \frac{\delta \mathcal{H}}{\delta d_b} \{d_b,S_a\}\\\nonumber
&=&
  \frac{\delta \mathcal{H}}{\delta {\bf S}} \times {\bf S}
+ \frac{\delta \mathcal{H}}{\delta {\bf d}} \times {\bf d},\\
\label{eq:ham_eq0b}
\dot d_a
= \{\mathcal{H}, d_a\}
&=&
  \frac{\delta \mathcal{H}}{\delta S_b} \{S_b,d_a\}
+ \frac{\delta \mathcal{H}}{\delta d_b} \{d_b,d_a\}\\\nonumber
&=&
  \frac{\delta \mathcal{H}}{\delta {\bf S}} \times  {\bf d}.
\end{eqnarray}
Using these equations one can show that $\frac{d}{dt}({\bf d\cdot S}) = 0$ and
thus the value $({\bf d\cdot S})$ is an integral of motion.

Derivatives of the Hamiltonian are:
\begin{eqnarray}
\label{eq:der1}
\frac{\delta \mathcal{H}}{\delta S_a}
&=&
- \gamma H_a
+ \frac{\gamma^2}{\chi_\perp}
\left[S_a + \delta\ ({\bf d} \cdot {\bf S}) d_a\right],\\
\label{eq:der2}
\frac{\delta \mathcal{H}}{\delta d_a}
&=&
\frac{\delta\ \gamma^2}{\chi_\perp}
({\bf d} \cdot {\bf S}) S_a\\\nonumber
&+& 4g_D\Delta^2\ ({\bf d} \cdot {\bf l}) l_a
- K_{jk}\Delta^2\ (\nabla_j\nabla_k d_a).
\end{eqnarray}
Substituting (\ref{eq:der1}), (\ref{eq:der2}), and~(\ref{eq:brackets_d})
into equations~(\ref{eq:ham_eq0a}-\ref{eq:ham_eq0b}) one has:
\begin{eqnarray}
{\bf\dot S}
&=&
[{\bf S} \times \gamma {\bf H}] \\\nonumber
&+&4g_D\Delta^2\ ({\bf d} \cdot {\bf l}) [{\bf l \times d}]
- K_{jk}\Delta^2\ [\nabla_j\nabla_k {\bf d} \times {\bf d}],
\\[2mm]
{\bf\dot d}
&=&
\gamma \left[{\bf d} \times
\left({\bf H} - \frac{\gamma}{\chi_\perp} {\bf S} \right)
\right].
\end{eqnarray}
Note that the anisotropy of susceptibility does not affect spin dynamics.


\subsection*{Linearized dynamics}

Consider small oscillations near the equilibrium:
\begin{equation}
{\bf S} = {\bf S^0} + {\bf\delta S}(t),\qquad
{\bf d} = {\bf d^0} + {\bf\delta d}(t)
\end{equation}

Linearize equations, differentiate the first one and exclude $\delta {\bf d}$. The result can be written as:
\begin{equation}
\delta\ddot S_a =
 [\delta {\bf\dot S} \times \gamma {\bf H}]_a + \Lambda_{ab}\ \delta S_b,
\end{equation}
where we introduce
\begin{eqnarray}\nonumber
\Lambda_{ab} &=& \Omega_P^2\ \big[
    ({\bf d^0} \cdot {\bf l})^2 \delta_{ab}
  - [{\bf l} \times {\bf d^0}]_a [{\bf l} \times {\bf d^0}]_b
  - ({\bf d^0} \cdot {\bf l}) d^0_a l_b
  \big]\\\nonumber
&+&c_{jk}^2\ \big[
  (\delta_{ab} - d^0_a d^0_b) \nabla_j\nabla_k
- 2 d^0_b (\nabla_j d^0_a) \nabla_k\\
&&\qquad +\ d^0_a (\nabla_j\nabla_k d^0_b)
- d^0_b (\nabla_j\nabla_k d^0_a)
\big],
\end{eqnarray}
\begin{equation}
\Omega_P^2 = 4g_D \frac{\Delta^2\gamma^2}{\chi_\perp},\qquad
c_{jk}^2 = K_{jk}\frac{\Delta^2\gamma^2}{\chi_\perp} = \Omega_P^2 \xi_{Djk}^2,
\end{equation}
and use the fact that $c_{jk} = c_{kj}$. Here~$\Omega_P$ is analog of Leggett
frequency, it determines NMR frequency shifts caused by spin-orbit interaction
and~$c_{jk}$ is anisotropic spin-wave velocity.

Consider~$H\parallel\hat z$ and look for a harmonic solution $\delta {\bf
S} = {\bf s}\exp(i\omega t)$. Then the equation can be written as
\begin{eqnarray}
-\omega^2 s_x &=&\Lambda_{xb}\ s_b + i \omega_L \omega\ s_y,\\\nonumber
-\omega^2 s_y &=&\Lambda_{yb}\ s_b - i \omega_L \omega\ s_x,\\\nonumber
-\omega^2 s_z &=&\Lambda_{zb}\ s_b
\end{eqnarray}
%
%


In high field (comparing with dipolar and gradient effects) motion of the spin
is close to a Larmor precession with frequency $\omega \approx \omega_L = \gamma H$ and $\Lambda \ll \omega_L^2$.
One can separate equations by putting~$s_y$ from the second equation to the first one and vise versa
and neglecting small terms. We get the same equations for~$s_x$ and~$s_y$. This can be written as a single
equation for a complex coordinate~$s_+ = (s_x + i s_y)/\sqrt{2}$:
\begin{equation}\label{eq:seq0}
(\omega_L^2-\omega^2) s_+ =  i (\Lambda_{xy}-\Lambda_{yx}) s_+  + (\Lambda_{xx} + \Lambda_{yy}) s_+
\end{equation}

In high field~$\bf d^0$ is perpendicular to the field and
we can use angles~(\ref{eq:angles}) with~$\beta_n=\pi/2$. Then
\begin{eqnarray}\nonumber
\Lambda_{xx} + \Lambda_{yy} &=&
\Omega_P^2\ \big[ (1+\sin^2\alpha )\sin^2\mu - 1 \big]
+c_{jk}^2\ \nabla_j\nabla_k
\\
\Lambda_{xy}-\Lambda_{yx} &=&
- \frac12 \Omega_P^2\ \sin 2\alpha\ \sin^2\mu \\\nonumber
&&¥+ 2 c_{jk}^2\ \big[
  (\nabla_j \nabla_k \alpha)
+ (\nabla_j \alpha) \nabla_k
\big].
\end{eqnarray}
Substituting this into~(\ref{eq:seq0}) and using~(\ref{eq:eq_alpha}) we have
\begin{eqnarray}\label{eq:spinwaves}
&&(\omega^2-\omega_L^2) s_+
= \Omega_P^2\ \Big\{\cos^2\mu - \sin^2\alpha\sin^2\mu\Big\}\ s_+
\\\nonumber
&&- c_{jk}^2\ \Big\{
  \nabla_j\nabla_k
+ i \big[
  (\nabla_j \nabla_k \alpha)
+ 2 (\nabla_j \alpha) \nabla_k
\big]
\Big\}\ s_+.
\end{eqnarray}
One can rewrite the equation in the form:
\begin{eqnarray}\label{eq:ab_spinwaves}
(\omega^2-\omega_L^2) s_+
&=& \Omega_P^2\ \Big\{ \cos^2\mu - \sin^2\alpha\sin^2\mu\Big\}\ s_+
\\\nonumber
&-& c_{jk}^2\ \Big\{
-\left( \frac{\nabla}{i} + \nabla\alpha \right)^2_{jk} + (\nabla\alpha)^2_{jk}
\Big\}\ s_+.
\end{eqnarray}
where we use notation $(X)^2_{jk} = X_j X_k$. This is similar to the
equation of motion of a charged particle in a magnetic field with a
vector potential ${\bf A} = \nabla\alpha$. The magnetic
field~$\nabla\times{\bf A}$ is zero everywhere except half-quantum vortex
cores but it affects the motion of the spin wave because of Aharonov-Bohm
effect~\cite{aharonov_bohm}. This effect for half-quantum vortices
in~$^3$He-A is discussed in~\cite{volovik_salomaa}. NMR and spin dynamics
of half-quantum vortices He-A are calculated in~\cite{hu_maki}.

For numerical calculations it is useful to make a substitution~$\bar s_+
= s_+ \exp(i\alpha)$. Then the equation for~$\bar s_+$ contains no
imaginary terms:
\begin{eqnarray}\label{eq:real_spinwaves}
(\omega^2-\omega_L^2) \bar s_+
&=& \Omega_P^2\ \Big\{\cos^2\mu - \sin^2\alpha\sin^2\mu\Big\}\ \bar s_+
\\\nonumber
&-& c_{jk}^2\ \Big\{ \nabla_j\nabla_k
+ (\nabla_j \alpha)(\nabla_k \alpha)
\Big\}\ \bar s_+,
\end{eqnarray}
The inverse transformation is needed if one need to calculate the actual
distribution of magnetization.


\image{h}{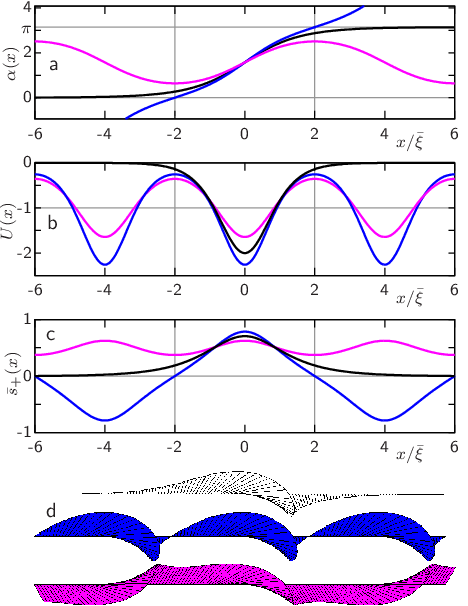}{Fig.~3. An example of the calculated texture and the spin wave
in 1D soliton structures. Black curves correspond to a single soliton,
blue and purple ones correspond to periodic structures with same and
alternating soliton orientations and periods $D=4\bar\xi$.
{\bf (a)} Texture,~$\alpha(x)$.
{\bf (b)} Potential for a real-value wave~$\bar s_+$. Energy levels
for all three textures are the same, $\lambda=-1$.
{\bf (c)} The real-value wave~$\bar s_+$.
{\bf (d)} Distribution of the amplitude and phase of the actual
magnetization~$s_+ = \bar s_+ \exp(-i\alpha)$. In all three cases the
total magnetization~$|\int s_+\,dx|$ is non-zero.
}
\imagew{t}{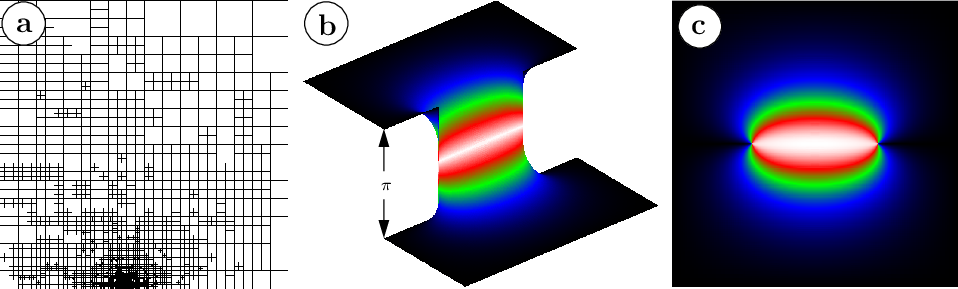}{Fig.~4. An example of the calculated texture and the spin wave
in the soliton between two half-quantum vortices.
{\bf (a)} The calculation grid made of 4696 cells covers one-forth of
the whole area $(8\times8)\,\bar\xi$ with two vortices separated by $D=3.5\,\bar\xi$.
Density of the grid is chosen according with gradients of the texture,
it is higher near vortices.
{\bf (b)} Calculated value of $\alpha$. One can see a smooth rotation by
$\pi$  in the soliton between vortices and $\pi$ jump on the other side
of vortices where phase also changes by~$\pi$.
{\bf (c)} The calculated real-value wave~$\bar s_+$.
}

\subsection*{NMR in the uniform texture and in the $d$-soliton}
To obtain frequency of the uniform NMR in the uniform texture we put $\alpha=0$
in~(\ref{eq:real_spinwaves}). Then the frequency is
\begin{equation}\label{eq:nmr_uniform}
\omega_u = \sqrt{\omega_L^2 + \Omega_P^2\ \cos^2\mu}.
\end{equation}
This formula can be used to measure~$\Omega_P$.

To find the spin wave, localized in the single $d$-soliton we use~(\ref{eq:real_spinwaves}) and
the soliton equation~(\ref{eq:alpha_sol}) for the distribution of~$\alpha$. This gives us
\begin{equation}\label{eq:1dsol}
\bar s_+ = \cosh^{-1} (x/\bar\xi),
\end{equation}
where as in~(\ref{eq:alpha_sol}) the value of~$\bar\xi$ depends on
the domain wall orientation. The frequency is
\begin{equation}\label{eq:nmr_soliton}
\omega_s = \sqrt{\omega_L^2 + \Omega_P^2\cos2\mu}.
\end{equation}
On NMR experiments two peaks are observed, one from the uniform texture
and another from waves localized in solitons. The difference between peak frequencies is
\begin{equation}
\delta\omega \approx \frac{\Omega_P^2}{2\omega}\sin^2\mu
\end{equation}
Intensity of the soliton peak (for a uniform rf-field) is proportional to
the oscillator strength~(\cite{hu_maki}), the ratio
\begin{equation}
I^M = \frac{\left|\int_V s_+\right|^2}{\int_V |s_+|^2}.
\end{equation}
This ratio also connects the total transverse magnetization, measured in
NMR experiments $M_\perp = \gamma\int_V s_+$ and energy stored in the
wave~$E = \gamma^2/2\chi_\perp \int_V |s_+|^2$. Ratio $I^M$ has a
dimension of volume. For a localized wave it is approximately equal to the volume
occupied by the wave. In one-dimensional case $I^M$ has a dimension of
length; for a single soliton~(\ref{eq:1dsol}) $I^M = 2\bar\xi$.

\subsection*{Numerical study of soliton structures}

For understanding results of real NMR experiments it is important to study
how various effects can change the frequency of the wave localized in the
soliton. We do it numerically in one and two-dimentional cases. Using
coordinates in units of~$\bar\xi^2$ one can write the
equation~(\ref{eq:eq_alpha}) for the texture as
\begin{equation}\label{eq:num_text}
\nabla^2\alpha = \frac12\sin2\alpha,
\end{equation}
and equation~(\ref{eq:real_spinwaves}) for the real-value waves as:
\begin{equation}\label{eq:num_wave}
- \nabla^2\ \bar s_+
+ U(x)\ \bar s_+
 = \lambda\ \bar s_+,
\end{equation}
where potential $U(x)=-(\nabla\alpha)^2 - \sin^2\alpha$ and
\begin{equation}
\lambda = \frac{\omega^2-\omega_L^2 - \Omega_P^2 \cos^2\mu}{\Omega_P^2 \sin^2\mu}
= -\frac{\omega^2-\omega_u^2}{\omega_s^2-\omega_u^2}.
\end{equation}
In the case of a single soliton $\omega=\omega_s$ and~$\lambda=-1$.

Using the equation~(\ref{eq:num_text}) we can numerically calculate
distribution of~$\alpha$. Then, using equation~(\ref{eq:num_wave}) we can
calculate eigenvalues~$\lambda$.

First consider a 1D periodic structure of parallel
solitons, located at some distance $D$ from each other. Solitons have an
orientation (direction of $\nabla\alpha$), and two simplest structures
which we study are sequences of solitons with same and alternating
orientations.

The solution for this problem is shown on Fig.~\ref{image:per_sol}.
Parameter $\lambda$ for both periodic structures has the same value~$-1$
as for the single soliton.

Let's also study an effect of a finite-length soliton.
Consider a two-dimensional problem with two half-quantum vortices
parallel to the~$l$ vector. Distance between vortices is~$D$. The same
equations~(\ref{eq:num_text}) and~(\ref{eq:num_wave}) are solved
numerically in 2D space using {\it deal.II} library~\cite{dealII}. The
code is available in~\cite{dealIIprog}. An example of the calculation is
presented on Fig.~\ref{image:calc_text1}.

\image{h}{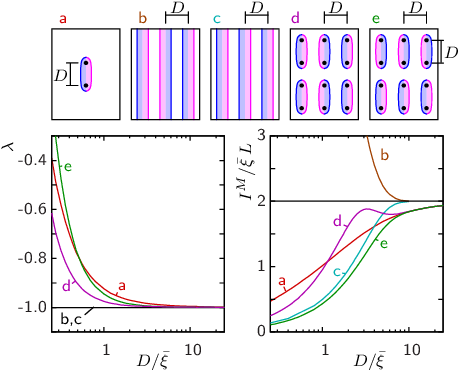}{Fig.~5. Calculated values of $\lambda$ and $I^M/\bar\xi L$
for various soliton structures. $L$ is the total soliton length.
}

Near a half-quantum vortex, at a distance much smaller
then~$\bar\xi$, the textural angle~$\alpha \approx
\varphi/2+\mbox{const.}$, where~$\varphi$ is the azimuthal coordinate.
One can see that the potential in~(\ref{eq:num_wave}) is $U(x) \approx
(\nabla\alpha)^2 \approx 1/4r^2$ (where $r$ is distance from the vortex
core). The real-value wave $\bar s_+$ can not fall into this hole because
of Aharonov-Bohm effect: it should be zero along some radial direction to
allow a smooth~$s_+$ distribution. The symmetry reasons tell, that in the
case of two vortices with a soliton the wave is zero on the line
connecting vortices outside them. The corresponding solution of the wave
equation is $\bar s_+
\approx \cos(\phi/2 + \mbox{const.})$, this kind of discontinuity is
clearly seen on the calculated wave near vortices.

On Fig.~\ref{image:num_res} calculated values of $\lambda$ and
$I^M/\bar\xi L$ ($L$ is the total soliton length) are plotted as
a function of some structure dimension $D/\bar\xi$. There are five
structures which are shown on the upper part of the figure: a single
soliton with a finite length~$D$; A periodic structures of infinite
solitons with the period~$D$ and same or alternating soliton
orientations; the combination of both effects, periodic structures of
finite solitons with equal length and period (this corresponds to a
square lattice of vortices). For large~$D$ all curves come to the values
for a single infinite soliton: $\lambda=-1$, $I^M=2\,\bar\xi L$. A
noticeable deviation of~$\lambda$ from the asymptotic value appears only
at high vortex densities, when the inter-vortex distance~$D$ is
comparable with~$\bar\xi$.

\subsection*{Conclusion}

Theoretical study of the texture and spin dynamics of the polar phase
is done. We start with the polar phase order parameter and the energy
with magnetic, opin-orbit and gradient terms. The order parameter
contains both orbital and spin anisotropy axes. The first one is fixed by
the aerogel strands, and the second can move in the plane perpendicular
to the applied magnetic field. Thus we have only one variable for the
texture, an angle~$\alpha$.  By minimizing the energy we get an equation
for the equilibrium texture. It has a form of a static sine-Gordon
equation for the angle~$\alpha$. We discuss possible topological defects
in this texture: $d$-solitons, half- and one-quantum vortices.
Characteristic length scale of the texture strongly depends on the
angle~$\mu$ between the magnetic field and aerogel strands. By rotating the
magnetic field one can vary it from a~$\xi_D \approx 20$~$\mu$m to infinity.

Numerical simulations of various textures with half-quantum vortices are
done. It is shown how interaction between vortices can change the NMR frequency.

\subsection*{Acknowledgements}
I thank G.E. Volovik for useful discussions. This work has been
supported in part by the Academy of Finland (project no. 284594).

\end{document}